\documentclass[letterpaper, preprint, paper,11pt]{AAS}	% for preprint proceedings

\usepackage{bm}
\usepackage{amssymb}
\usepackage{amsmath}
\usepackage{subfigure}
\usepackage[colorlinks=true, pdfstartview=FitV, linkcolor=black, citecolor= black, urlcolor= black]{hyperref}
\usepackage{overcite}
\usepackage{footnpag}			      	% make footnote symbols restart on each page

\usepackage[normalem]{ulem}
\usepackage{soul}
\usepackage{cancel}
\usepackage{xcolor}

\PaperNumber{24-295}

\begin{document}

% \title{MANUSCRIPT TITLE (UP TO 6 INCHES IN WIDTH AND CENTERED, 14 POINT BOLD FONT, MAJUSCULE)}
\title{Constrained Attitude Reorientation Maneuver of Spacecraft based on Path Planning}

\author{Daegyun Choi\thanks{Postdoctoral Researcher, Department of Aerospace Engineering \& Engineering Mechanics, University of Cincinnati, Cincinnati, OH 45221, USA.},  
Donghoon Kim\thanks{Assistant Professor, Department of Aerospace Engineering \& Engineering Mechanics, University of Cincinnati, Cincinnati, OH 45221, USA.}, \ and 
Henzeh Leeghim\thanks{Professor, Department of Aerospace Engineering, Chosun University, Gwangju 61452, Republic of Korea.}
% ,
% Jaehyun Jin\thanks{Professor, Department of Aerospace Engineering, Sunchon National University,
% Suncheon, Jeonnam 57922, Republic of Korea.}
}

\maketitle{} 		

\begin{abstract}
Spacecraft attitude control becomes challenging when light-sensitive instruments must avoid bright objects. This work proposes a path-planning-based spacecraft attitude reorientation maneuver considering these constraints to reach the desired attitude. Constraint-compliant attitude trajectories are computed using a fuzzy enhanced potential field approach in the attitude space. Based on the path-planning output, the control input is determined by considering the dynamics and kinematics of the spacecraft to satisfy dynamic constraints in the time domain. Numerical simulation results demonstrate successful reorientation maneuvers without violating the constraints.
\end{abstract}

\section{Introduction}

Since the beginning of the spaceflight era, numerous Earth observation spacecraft have been launched to monitor the Earth’s land, oceans, atmosphere, and carbon cycle in real-time, transmitting information to ground stations \cite{EarthObservation}. 
Among these, Earth-imaging spacecraft, which use optical sensors to capture images, account for the largest number \cite{ImagingSatellite}. A key process for these spacecraft is adjusting their attitude to align the sensor's boresight vectors with the desired direction%. In fact, such reorientation maneuvers become challenging missions when there are constraints on the spacecraft's
, a task complicated by orientation constraints. These constraints can be categorized as keep-out and keep-in constraints \cite{Kjllberg2016}. Keep-out constraints ensure that light-sensitive instruments, such as optical sensors or star trackers, do not point at bright celestial objects like the Sun or the Moon \cite{Kjllberg2016}. These instruments have a cone-shaped field of view (FOV), and the bright object must be kept out of this FOV. % On the other hand, k
Keep-in constraints require some sensors to face a specific celestial object or direction within a certain angular distance \cite{Yang2021}. These include solar panels maintaining a specific incident angle to the sunlight for steady power generation or sun sensors needing to keep the Sun within their FOV.
% \hl{Includes citations where not addressed.}

%To successfully complete missions with such constraints, many r
Researchers have studied various methods to manage these constrained reorientation maneuvers. Some have proposed incorporating potential functions into the control law, where the potential value increases near keep-out constraints or the boundary of keep-in constraints and decreases as the spacecraft approaches the desired attitude \cite{PF2024_1,PF2024_2,PF2021_1}. These approaches are robust and easy to implement but can struggle with complex or non-convex geometries and may get trapped in local minima \cite{PF_disadv, PF_disadv2}. Other approaches involve path-planning, such as using probabilistic roadmaps for a randomized search \cite{prob_roadmap} or discretizing the attitude space to find a route from the initial to the desired attitude \cite{PP2015,PP2017,PP2020,PP2022}.
While probabilistic roadmaps provide smooth trajectories, their randomness does not guarantee convergence. 
Discretized approaches offer a set of attitude waypoints, but these typically require additional smoothing \cite{PP2022}. 
Some researchers have aimed to find an optimal attitude trajectory to minimize control effort, but these methods generally require off-line computation \cite{PP2020,PP2022}.

This work focuses on real-time, constraint-compliant reorientation maneuvers from an initial to a desired attitude while avoiding keep-out constraints. Initially, the given keep-out constraints are represented in the attitude space using modified Rodrigues parameters (MRPs) for a minimal three-parameter coordinate set \cite{PP2022}.
The appropriate attitude route is then determined using the fuzzy enhanced potential field (FEPF) approach \cite{Choi2021Robotica}, which offers intuitive principles and smooth trajectories in real-time without local minima \cite{Choi2021AP}. The FEPF avoids the need for additional smoothing processes and adaptively searches for near-optimal routes based on linguistic variable-based inputs and outputs and the if-then rules. 
% Although the attitude trajectory in the MRP space is obtained from the previous step, this may violate the dynamic constraints.
% , such as the limitations of the angular rates and control torque. 
% Hence, 

Originally developed for unmanned aerial vehicles, the FEPF parameters are scaled down and revised for spacecraft use. 
Based on the FEPF-derived attitude information, the motion planning process considers the spacecraft's dynamics to determine a dynamics-compliant attitude trajectory. 
% This is the mapping process from the nondimensional attitude description to the actual time domain with the satisfaction of the dynamic constraints. 
%Through this process, the 
This results in proper angular velocity and control input trajectories in the time domain. The proposed approach's performance is validated through numerical simulations using multiple scenarios.

\section{Preliminaries}
\begin{figure}[hbpt]
    \centering
    \includegraphics[width=0.6\linewidth]{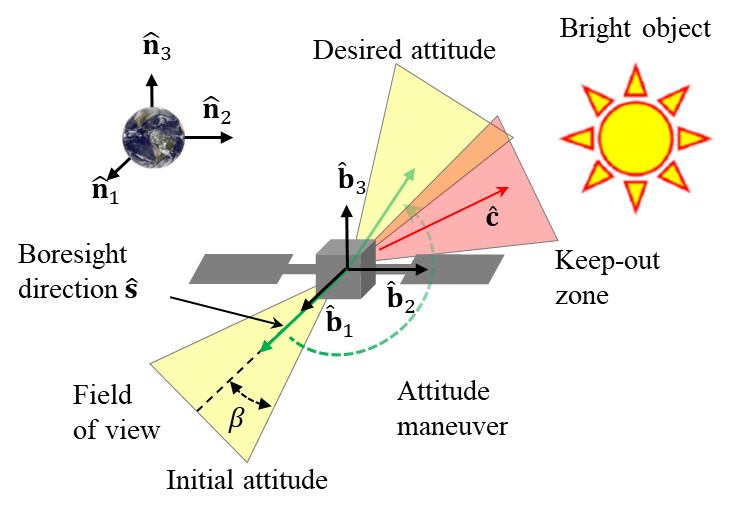}
    \caption{Conceptual Representation of a Constrained Attitude Maneuver}
    \label{fig:concept_const}
\end{figure}

%\subsection{Problem Description}
% \begin{figure}[hbpt]
%     \centering
%     \includegraphics[width=0.6\linewidth]{files/fig_concept2.png}
%     \caption{Conceptual representation of a constrained attitude maneuver}
%     \label{fig:concept_const}
% \end{figure}
This work studies the spacecraft attitude reorientation maneuver from the initial to the desired attitude in the presence of constraints, as shown in Figure~\ref{fig:concept_const}. 
The constraints are defined as keep-out zones, where a boresight vector and a FOV of a light-sensitive instrument (green arrow and yellow region) should be kept away from bright objects (red arrow) during maneuver. This work aims to find the constraint-compliant attitude trajectory in the non-dimensional attitude space while adhering to the dynamics and kinematics of spacecraft.

\subsection{Dynamics and Kinematics% of Spacecraft
}
The attitude of spacecraft is defined by the rotation of the body-fixed frame $\{O_B: \hat{\bf b}_1,\hat{\bf b}_2,\hat{\bf b}_3\}$ with respect to the inertial frame $\{O_N: \hat{\bf n}_1,\hat{\bf n}_2,\hat{\bf n}_3\}$. This work considers the MRPs because they provide a minimal three-parameter coordinate set \cite{PP2022,Schaub2003analytical}. Based on the principal rotation vector and angle, $\hat{\bf e}\in\mathbb{R}^3$ and $\Phi\in\mathbb{R}$, the MRP set is defined as \cite{Schaub2003analytical}
\begin{equation}
    \bm\sigma =
    \begin{bmatrix}
        \sigma_1 \\ \sigma_2 \\ \sigma_3
    \end{bmatrix}
    =\frac{1}{1+\cos{\frac{\Phi}{2}}}
    \begin{bmatrix} e_1 \sin{\frac{\Phi}{2}} \\ e_2 \sin{\frac{\Phi}{2}} \\ e_3 \sin{\frac{\Phi}{2}}
    \end{bmatrix}
    =\hat{\bf e}\tan{\frac{\Phi}{4}}.
\end{equation}
For every principal rotation set $(\hat{\bf e},\Phi)$, there exists a shadow set $(\hat{\bf e},\Phi')$, where $\Phi'=\Phi-2\pi$ that represents the same attitude. The shadow set also defines the shadow MRP set expressed by
\begin{equation}
    \sigma_i^S = \frac{e_i \sin{\frac{\Phi'}{2}}}{1+\cos{\frac{\Phi'}{2}}} = \frac{-\sigma_i}{\bm\sigma^\mathrm{T}\bm\sigma}{,}\  (\text{for }i=1,2,~\text{and}~3).
\end{equation}
Further explanations for two MRP sets will be described in the next section describing the MRP space.

The differential kinematic equation of the MRPs is expressed \cite{Schaub2003analytical}
\begin{equation}\label{eq:sigmadot}
    \dot{\bm\sigma} = \frac{1}{4}B \ {}^B\bm\omega = \frac{1}{4}\left[(1-\bm\sigma^\mathrm{T}\bm\sigma)I_{3\times 3} + 2[\bm\sigma^{\times}] + 2\bm\sigma\bm\sigma^\mathrm{T}\right] \ {}^B\bm\omega,
\end{equation}
% \begin{equation}
%     B = (1-\bm\sigma^\mathrm{T}\bm\sigma)I_{3\times 3} + 2[\bm\sigma^{\times}] + 2\bm\sigma\bm\sigma^\mathrm{T},
% \end{equation}
where $[\bm\eta^\times]\in\mathbb{R}^{3\times 3}$ is the skew-symmetric matrix composed of each element of a generic vector $\bm\eta\in\mathbb{R}^3$ and ${}^B\bm\omega\in\mathbb{R}^3$ is the angular velocity of spacecraft with respect to the inertial frame ($N$), expressed in the body frame ($B$). For the notation simplicity, the superscript $B$ will be omitted. 

Euler's equation describing the rotational motion of spacecraft is expressed as \cite{Schaub2003analytical}
\begin{equation}\label{eq:dynamics}
    J\dot{\bm\omega} = -[\bm\omega^{\times}] J \bm\omega + {\bf u},
\end{equation}
where $J\in\mathbb{R}^{3\times 3}$ is the spacecraft inertia, and ${\bf u}\in\mathbb{R}^3$ is the control input.

\subsection{Constraint Representation}\label{sec:constraint}

The constraints of spacecraft are defined by the boresight vector and the FOV of the light-sensitive instrument relative to the direction vectors of bright objects. The bright objects should not be within the FOV of the light-sensitive instrument during reorientation maneuvers, as illustrated in Figure~\ref{fig:concept_const}. This means that the body-fixed direction of the $i$-th light-sensitive instrument ($\hat{\bf s}_i\in\mathbb{R}^3$) must be away from the red region, defined by the direction of the $j$-th bright object (${}^N\hat{\bf c}_j\in\mathbb{R}^3$). Therefore, the following constraint must be satisfied:
\begin{equation}\label{eq:constraint}
    \hat{\bf s}_i^\mathrm{T} C_{BN} \ {}^N\hat{\bf c}_j < \cos\beta_i \ (\text{for }i=1,..., i_s, \text{ and } j=1, ..., j_c),
\end{equation}
where $\beta_i$ is the half-angle of the FOV of the $i$-th instrument, $C_{BN}$ is the direction cosine matrix that maps from the inertial frame to the body frame, $i_s$ is the number of instruments, and $j_c$ is the number of bright objects. 
These constraints are usually visualized as a cone shape in the unit sphere, with the vertex is located at the origin of the body frame, as shown in Figure~\ref{fig:concept_const}.

\section{Methodology}
The focus of this work is to find the constraint-compliant attitude trajectory for spacecraft to reorient the attitude to the desired attitude. 
% Since a path planning approach for the constrained attitude maneuver is employed, 
This work first introduces a workspace defined in the MRP space, composed of three MRP components, and converts the states, including the constraints, into the components in the MRP space. Then, a path-planning approach is applied to find the trajectory in the MRP space from the initial to the desired states while avoiding obstacles, which are the constraints defined in the MRP space. Based on the output obtained from the path-planning approach, the attitude and angular velocity can be computed using the kinematic and dynamic equations.

\subsection{State Conversion to the MRP space}\label{sec:mrp_space}

% This work aims to find the attitude trajectory without violoating constraints. For this reason, 
This work considers the MRP space as a workspace to employ the path-planning approach for finding the constraint-compliant attitude trajectory. The three components of the MRP set representing the spacecraft attitude can be visualized in three-dimensional (3D) Cartesian coordinate system as a vector. 
Note that any MRP set whose norm is greater than 1 ($||\bm\sigma|| > 1$) has a corresponding shadow MRP set whose norm is smaller than 1 ($||\bm\sigma^S|| < 1$). That is, for any MRP set represented by a point outside of a 3D unit sphere in the MRP space, there exists a shadow MRP set within the same unit sphere representing the same attitude. Therefore, any 
spacecraft attitude is defined within a unit sphere centered at the origin in the MRP space.

The constraints defined in the inertial space should be converted into the MRP space.
Groups of MRP vectors that violate the constraints defined in Eq. \eqref{eq:constraint} are represented as areas with volume and are regarded as obstacles in the MRP space. The constrained regions are illustrated in the simulation study section. This way, the constrained attitude reorientation maneuver problem is transformed into finding the MRP vector's 3D trajectory in the MRP space while avoiding collision with obstacles. 

% For example, consider the boresight vector with the half angle of the FOV of 20 deg as ${\bf s}_1 = [1,0,0]^\mathrm{T}$ and the bright objects's direction vectors as ${\bf c}_1 = [0, -0.981, 0.196]^\mathrm{T}$, ${\bf c}_2 = [-1, -0, 0]^\mathrm{T}$, and ${\bf c}_3 = [0.958, 0, 0.287]^\mathrm{T}$. Each keep-out zone in the inertial space is expressed in Figure~\ref{fig:const_in}, and the converted keep-out zone in the MRP space is described as Figure~\ref{fig:const_mrp}.
% \begin{figure}[htbp]
%     \centering
%     \subfigure[Keep-out zone in the inertial space]{\label{fig:const_in}
%     \includegraphics[width=0.48\columnwidth]{files/fig_constraint.png}}
%     \subfigure[Keep-out region in the MRP space]{\label{fig:const_mrp}   
%     \includegraphics[width=0.48\columnwidth]{files/fig_constraint_mrp.png}}
%     \caption{Keep-out constraints ($\beta=20$ deg) \blue{[Revise circles to cones in the left figure]}}
%     \label{fig:const}
% \end{figure}
% That is, the colored regions (keep-out zone) are regarded as obstacles to be avoided while find the path from the initial to the desired MRP vectors.

\subsection{Constraint-Compliant MRP Vector Generation}
Given the initial and desired MRP vectors ($\bm\sigma_0\in\mathbb{R}^3$ and $\bm\sigma_d\in\mathbb{R}^3$) with obstacles (keep-out zones) defined in the MRP space, the next step is to find the MRP vectors to reach the desired attitude while avoiding the keep-out zones. This work employs a path-planning approach, FEPF \cite{Choi2021Robotica}. This method, developed by the authors, was designed and trained to find a near-optimal collision-free trajectory in the inertial space in terms of the path length based on the concept of the artificial potential field. 
% By combining the enhanced potential field and two fuzzy inference systems (FISs), the FEPF approach does not encounter local minima and goal non reachable obstacles nearby issues, which are common issues in the path planning field. 
At the current state ($\bm\sigma\in\mathbb{R}^3$) in the MRP space, the FEPF outputs a vector ($\dot{\bm\sigma}_t\in\mathbb{R}^3$), which is regarded as the time derivative of the state, to determine the next state at the next time step. The output vector is defined by \cite{Choi2021Robotica}
\begin{equation}
    \dot{\bm\sigma}_t = m_a \hat{\bm\sigma}_a(\bm\sigma, \bm\sigma_d) + m_r \hat{\bm\sigma}_r(\bm\sigma),
\end{equation}
where $\hat{\bm\sigma}_a\in\mathbb{R}^3$ and $\hat{\bm\sigma}_r\in\mathbb{R}^3$ are the attractive (toward the desired state) and repulsive (away from the keep-out zones) potential field directions, respectively. Also, $m_a\in\mathbb{R}$ and $m_r\in\mathbb{R}$ are the magnitude of each potential field, respectively. Two fuzzy inference systems (FISs) in the FEPF approach determine $m_a$ and $m_r$, while the enhanced potential field (EPF) provides the directions of each potential field. 
% \hl{Lack of the information about the EPF.}
%\blue{It is important to note that the EPF is an improved version of potential field-based path-planning approaches that resolves the well-known issues, such as local minima and goal non-reachable obstacles nearby 
The EPF represents an advanced version of traditional potential field-based path-planning approaches, addressing common issues such as local minima and obstacles that prevent reaching the goal \cite{Choi2020EPF2D, Choi2021AP, Choi2024RPO}.

For $m_a$ and $m_r$, since the workspace of the FEPF is different from the original application described in the authors' previous work \cite{Choi2021Robotica}, the input range of the two FISs should be revised, especially for the inputs for determining $m_a$ (the Euclidean distance between the current and desired states) and $m_r$ (the Euclidean distance between the current and the closest constrained region). The input range for this work is reduced to between 0 to 1 because the workspace is inside the unit sphere produced by three components of the MRP vector. 
The rules that map from the inputs to the output remain the same because the behavior for finding the collision-free trajectory is identical. Detailed information for the magnitude determination process is described in the reference \cite{Choi2021Robotica}.

For $\hat{\bm\sigma}_a$ and $\hat{\bm\sigma}_r$, the directions are defined in 3D space, although the EPF in the study \cite{Choi2021Robotica} is derived in 2D space. Therefore, for this work, the direction determination formula of the concept of the EPF in 3D space is employed as \cite{Choi2024RPO}
\begin{align}
    \hat{\bm\sigma_a} & = \frac{\bm\sigma_d-\bm\sigma}{||\bm\sigma_d-\bm\sigma||},\\
    \hat{\bm\sigma}_r & = -R\hat{\bm\sigma}_c,
\end{align}
where $-\hat{\bm\sigma}_c$ indicates the unit direction vector from the closest point between the current MRP vector $\bm\sigma$ and the keep-out zone to $\bm\sigma$, and $R\in\mathbb{R}^{3\times 3}$ is a rotation matrix defined as
\begin{equation}
    R = \begin{bmatrix}
        o_1^2 \kappa + \cos\gamma & o_1 o_2\kappa - o_3\sin\gamma & o_1o_3\kappa + o_2\sin\gamma\\
        o_2 o_1 \kappa + o_3 \sin\gamma & o_2^2 \kappa + \cos\gamma & o_2 o_3 \kappa - o_1 \sin\gamma \\
        o_3o_1\kappa - o_2\sin\gamma & o_3o_2\kappa +o_1\sin\gamma & o_3^2 \kappa+\cos\gamma
    \end{bmatrix}.
\end{equation}
Here, $\gamma\in\mathbb{R}$ is a rotation angle, $\kappa=1-\cos\gamma$, and $o_i$ for $i=1,2,~\text{and}~3$ is the element of a vector ($\hat{\bf o}$) defined by 
\begin{equation}
    \hat{\bf o} = \frac{[\hat{\bm\sigma}_c^\times]  \bm\sigma}{||\hat{\bm\sigma}_c|| ||\bm\sigma||}.
\end{equation}

Once the output vector $\dot{\bm\sigma}_t$ from the FEPF is obtained, the next MRP vector is computed using Eq. \eqref{eq:sigmadot} via %a 
numerical integration, and the obtained MRP vector will be the target MRP vector $\bm\sigma_t$ to track at the current time step to reach $\bm\sigma_d$ ultimately. Here, $\bm\sigma_t$ complies with the MRP kinematics in Eq. \eqref{eq:sigmadot} and the constraints in Eq. \eqref{eq:constraint}.

\subsection{Physics-Compliant States Propagation}
% When the scale-revised FISs of the FEPF are applied for the constrained attitude maneuver, this assumes that ${\bf p}_t$ is equal to $\dot{\bm\sigma}_\text{int}$ considering the aspect of the path planning.
Using the obtained $\bm\sigma_t$ and $\dot{\bm\sigma}_t$ from the previous step, one can obtain the physics-satisfied states, such as the torque, attitude, and angular velocity, based on the dynamic and kinematic equations. 
To comply with the dynamics, the control torque should be generated. This work considers a PD controller to produce the control input ${\bf u}\in\mathbb{R}^3$, defined as
\begin{equation}\label{eq:pdcontrol}
    {\bf u} = -K_p{\bm\sigma_e} - K_d{\bm\omega_e},
\end{equation}
where $K_p \in\mathbb{R}^{3\times 3}$ and $K_d \in\mathbb{R}^{3\times 3}$ are the control gain matrices, $\bm\sigma_e$  is the MRP error vector between the current and target MRP vector, and ${\bm\omega_e}$ is the angular velocity error between the current and target angular velocity. Similar to $\bm\sigma_t$, $\bm\omega_t$ is defined as the target angular velocity to track at the current time step and is determined by {the kinematic equation as follows:}
\begin{equation}\label{eq:angvel}
    \bm\omega_t = \frac{4}{(1+\bm\sigma_t^\mathrm{T}\bm\sigma_t)^2}B^\mathrm{T}\dot{\bm\sigma}_t.
\end{equation}
The output of the FEPF ($\dot{\bm\sigma}_t$) is utilized to determine $\bm\sigma_t$ and $\bm\omega_t$, which can determine the control input in Eq. \eqref{eq:pdcontrol}. Using the control input obtained, one can propagate the angular velocity complying with the dynamics using Eq. \eqref{eq:dynamics}. In addition, the kinematics-compliant MRP set can be obtained via the kinematic equation in Eq. \eqref{eq:sigmadot} using the current MRP set ($\bm\sigma$) and angular velocity ($\bm\omega$).

\section{Simulation Study}\label{sec:simulation}
%\subsection{Simulation Parameters}
To validate the performance of the proposed approach, numerical simulations are conducted for a spacecraft containing one light-sensitive instrument in the presence of three bright objects around the spacecraft. The simulation parameters are tabulated in Table \ref{tab:parameters}.
\begin{table}[htbp]
    \centering
    \caption{Simulation Parameters}
    \label{tab:parameters}
    \begin{tabular}{c|c}
    \hline
        Parameters & Values \\
        \hline
        Simulation time & 150 (s) \\
        Time interval ($\mathrm{d}t$) & 0.1 (s)\\
        Spacecraft inertia ($J$)& diag$([10,10,10]) $ (kgm$^2$)\\
        Initial MRP set ($\bm\sigma_0$) & Case 1: $[0,0,0.25]^\mathrm{T}$, Case 2: $[0.3,0.25,0]^\mathrm{T}$\\
        Desired MRP set ($\bm\sigma_d$) & Case 1: $[0,-0.2,-0.7]^\mathrm{T}$, Case 2: $[0.1,-0.3,0]^\mathrm{T}$\\
        Initial angular velocity ($\bm\omega_0$) & $[0,0,0]^\mathrm{T}$ (rad/s)\\
        Boresight vector ($\hat{\bf s}$) & $[1, 0, 0]^\mathrm{T}$\\
        Half angle of the FOV ($\beta$) & 20 (deg) \\
        Bright object 1 ($\hat{\bf c}_1$) & $[0, -0.981, -0.196]^\mathrm{T}$\\
        Bright object 2 ($\hat{\bf c}_2$) & $[-1, 0, 0]^\mathrm{T}$\\
        Bright object 3 ($\hat{\bf c}_3$) & $[0.958, 0, 0.287]^\mathrm{T}$\\
        Proportional gain ($K_p$) & 5.5 $J$ \\
        Derivative gain ($K_d$) & 3.1 $J$\\
        \hline
    \end{tabular}
\end{table}
% \begin{figure}[htbp]
%     \centering
%     \subfigure[Keep-out zone in the inertial space]{\label{fig:const_in}
%     \includegraphics[width=0.35\columnwidth]{files/fig_constraint.png}}
%     \subfigure[Keep-out region in the MRP space]{\label{fig:const_mrp}   
%     \includegraphics[width=0.35\columnwidth]{files/fig_constraint_mrp.png}}
%     \caption{Keep-out constraints \blue{[Revise circles to cones in the left figure and add the legend]}}
%     \label{fig:const}
% \end{figure}
\begin{figure}[htbp]
    \centering
    \subfigure[{Constraints} in the inertial space]{\label{fig:const_in}
    \includegraphics[width=0.4\columnwidth]{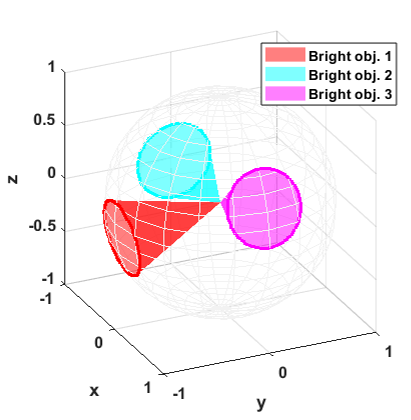}}
    \subfigure[{Constraints} in the MRP space]{\label{fig:const_mrp}   
    \includegraphics[width=0.4\columnwidth]{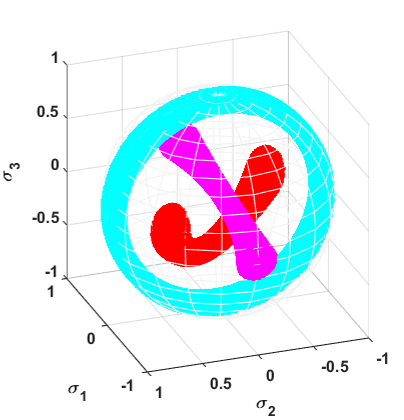}}
    \caption{{Keep-Out Constraints} 
    % \blue{[Revise the circles to cones and add the legend]}
    }
    \label{fig:const}
\end{figure}
\subsection{Visualization of the Keep-Out Zone in the Inertial and MRP Spaces}
The constraints related to bright objects defined in the inertial space are converted into the MRP space. Considering the parameters listed in Table \ref{tab:parameters}, each keep-out zone in the inertial space is expressed in Figure~\ref{fig:const_in}, and the converted keep-out zones in the MRP space are illustrated in Figure~\ref{fig:const_mrp}. 
% \begin{figure}[htbp]
%     \centering
%     \subfigure[{Constraints} in the inertial space]{\label{fig:const_in}
%     \includegraphics[width=0.4\columnwidth]{files/fig_constraint2.png}}
%     \subfigure[{Constraints} in the MRP space]{\label{fig:const_mrp}   
%     \includegraphics[width=0.4\columnwidth]{files/fig_constraint_mrp.png}}
%     \caption{{Keep-out constraints} 
%     % \blue{[Revise the circles to cones and add the legend]}
%     }
%     \label{fig:const}
% \end{figure}
The colored regions (keep-out zones) are regarded as obstacles to be avoided while finding the MRP vectors from the initial to the desired MRP vectors.

\subsection{Simulation Results}
\begin{figure}[htbp]
    \centering
    \subfigure[MRP vector trajectory]{\label{fig:3d1}
    \includegraphics[width=0.49\columnwidth]{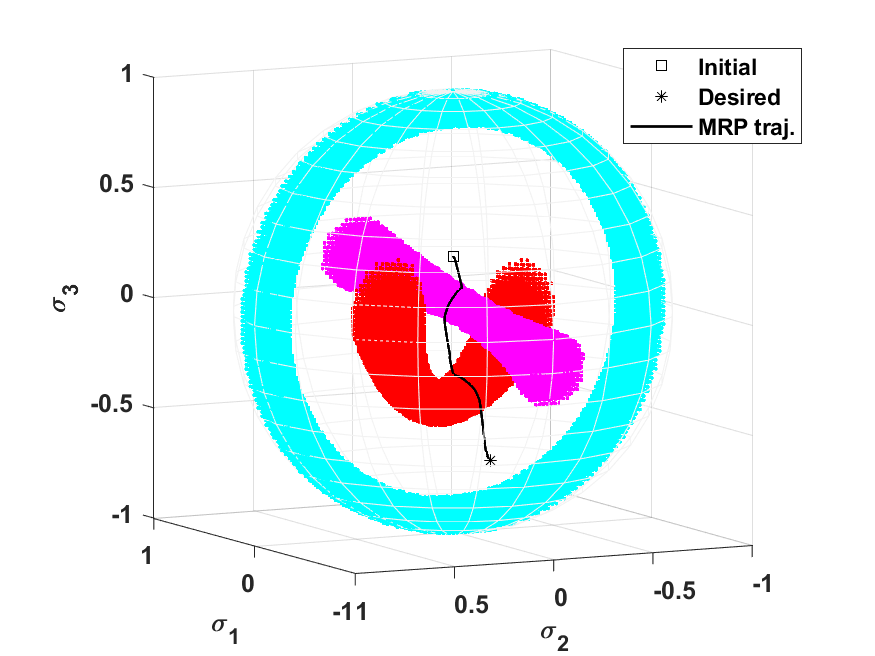}}
    \subfigure[Boresight vector trajectory]{\label{fig:2d1}   
    \includegraphics[width=0.49\columnwidth]{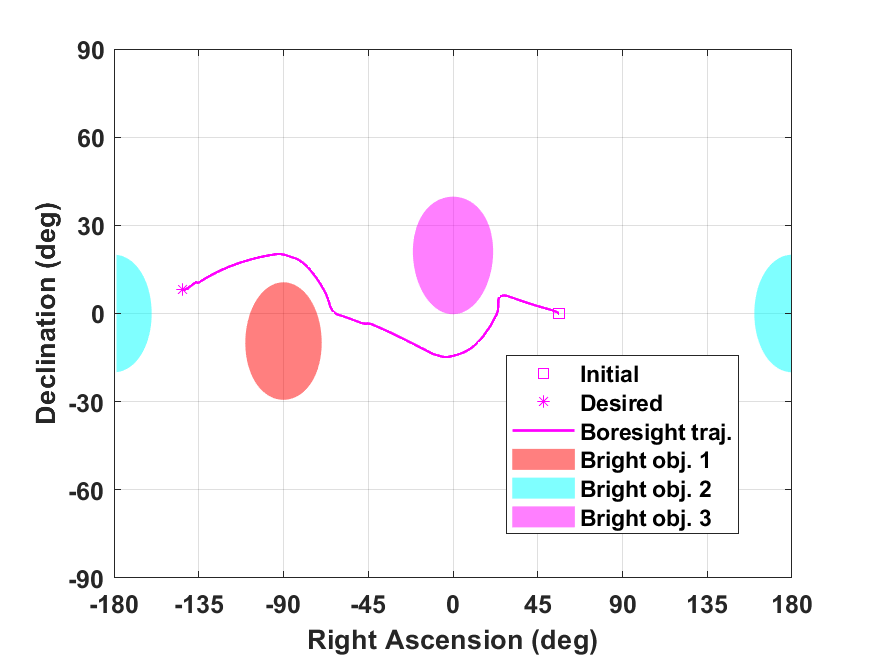}}
    \caption{MRP Vector Trajectory in the MRP Space and the Boresight Vector Trajectory in the Inertial Space (Case 1)}
    \label{fig:traj1}
\end{figure}

\begin{figure}[htbp]
    \centering
    \subfigure[MRPs]{\label{fig:mrp1}
    \includegraphics[width=0.46\columnwidth]{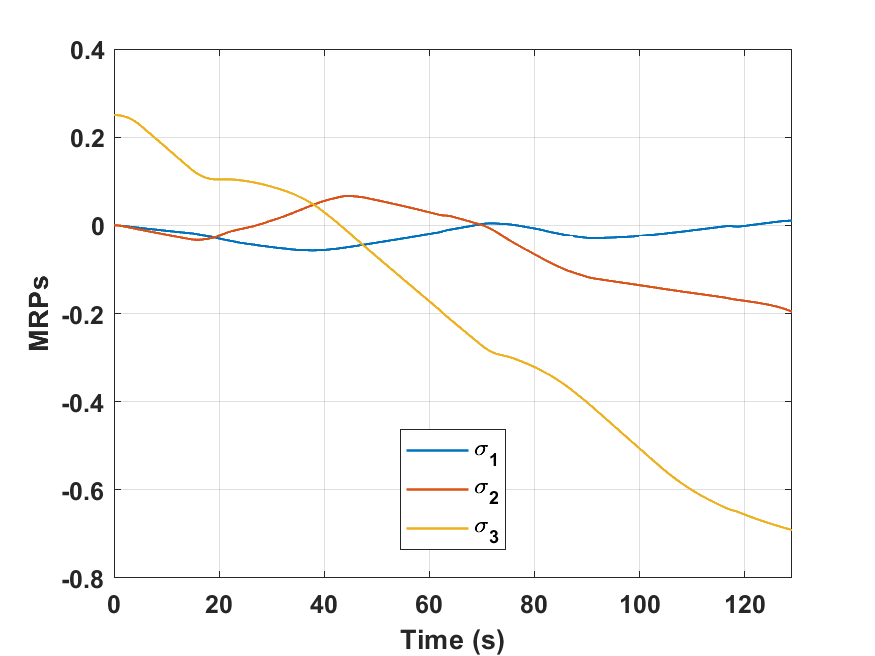}}
    \subfigure[Angular velocity]{\label{fig:angvel1}   
    \includegraphics[width=0.46\columnwidth]{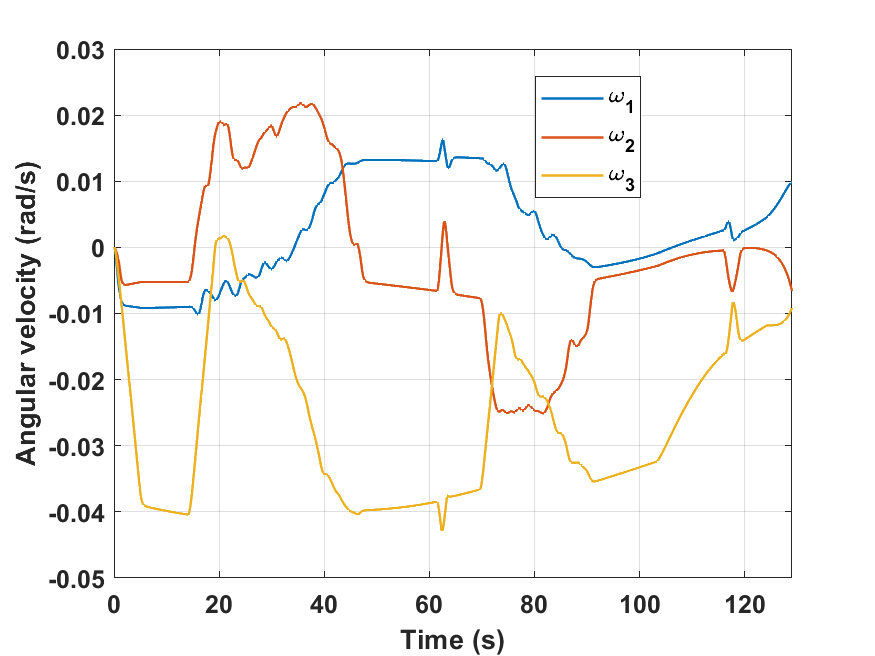}}
    \subfigure[Torque]{\label{fig:torque1}   
    \includegraphics[width=0.46\columnwidth]{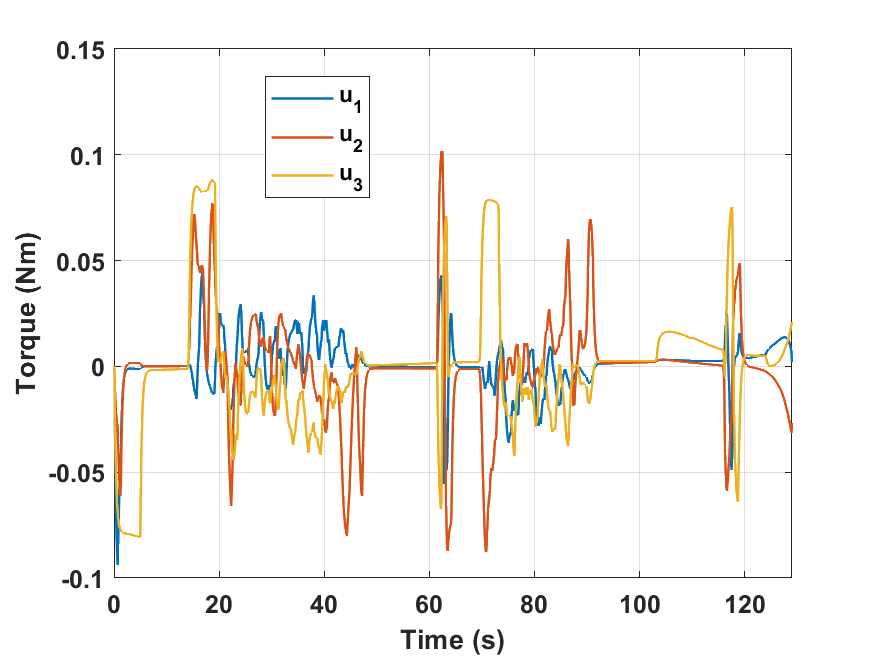}}
    \caption{Time History of the States (Case 1)}
    \label{fig:timehist1}
\end{figure}

In Case 1, the desired MRP set is given that the spacecraft's attitude change is dominantly in the third component of the MRP set, as confirmed by the MRP vector's trajectory in the MRP space shown in Figure \ref{fig:3d1}. A smooth MRP trajectory in the MRP space is obtained using the FEPF's output. The boresight trajectory, including the keep-out constraints in the inertial space, is visualized by projecting it onto a 2D plane composed of the right ascension and declination, as illustrated in Figure~\ref{fig:2d1}. 
In 2D space, the boresight vector also changes smoothly while avoiding the keep-out zones. The smooth MRP trajectories are also observed in the time history of the MRP set in Figure~\ref{fig:mrp1}. Although the given simulation time is 150 seconds, the maneuver is achieved in 129 seconds. On the other hand, the time histories of the angular velocity in Figure~\ref{fig:angvel1} and torque in Figure~\ref{fig:torque1} contain non-smooth trajectories. The fluctuations in the states are observed when the MRP vector is near the keep-out zones. In particular, the torque trajectory includes more peaks and fluctuations compared to the angular velocity. Also, the terminal values of the angular velocity and torque are non-zero. This is because the current work focuses on finding the shortest trajectory in the MRP space, subject to the constraints.
Smoothing of both the angular velocity and torque constraints, as well as terminal angular velocity constraints, will be considered in the future training of the FEPF. 

\begin{figure}[htbp]
    \centering
    \subfigure[MRP vector trajectory]{\label{fig:3d2}
    \includegraphics[width=0.49\columnwidth]{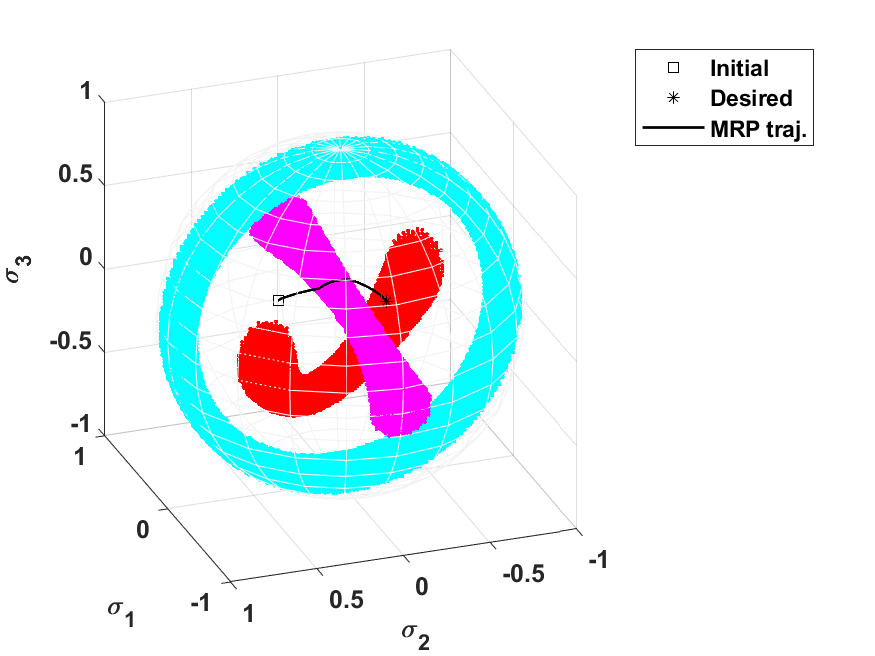}}
    \subfigure[Boresight vector trajectory]{\label{fig:2d2}   
    \includegraphics[width=0.49\columnwidth]{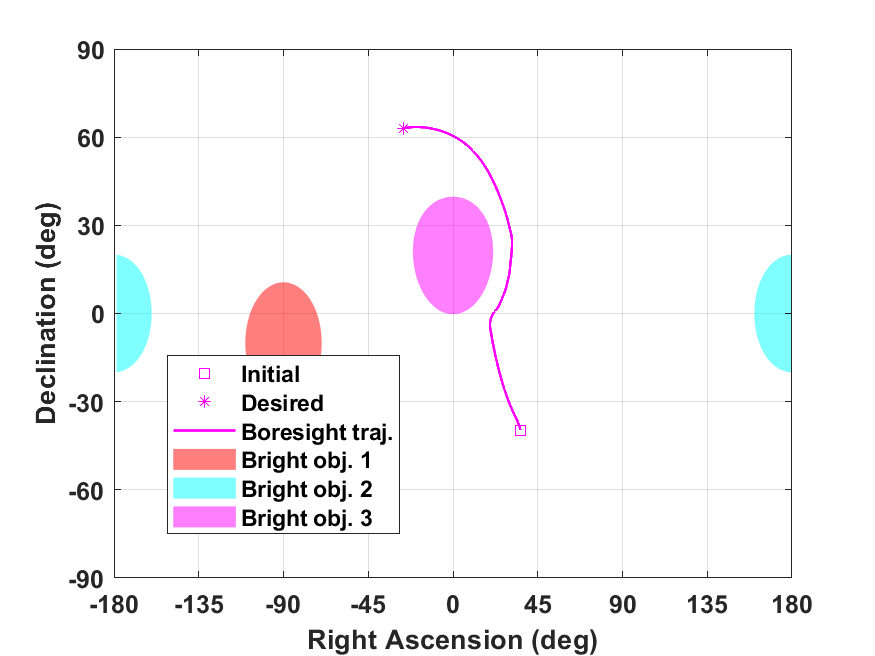}}
    \caption{MRP Vector Trajectory in the MRP Space and the Boresight Vector Trajectory in the Inertial Space (Case 2)}
    \label{fig:traj2}
\end{figure}

In Case 2, the dominant attitude change is in the second term in the MRP set. As shown in Figure \ref{fig:2d2}, a horizontal and smooth MRP trajectory is obtained. The boresight trajectory in the projected plane is also smooth and does not violate the constraints. Since the total MRP trajectory length in this case is shorter than in Case 1, the reorientation maneuver is completed in 72 seconds, earlier than in Case 1. 
Similar to Case 1, the time history of each component of the MRP set is very smooth. However, the angular velocity and torque trajectories contain the same issues, with several peaks while avoiding the keep-out zones. These undesired behaviors, especially for the angular velocity and torque, will be addressed in future work. 
\begin{figure}[t]
    \centering
    \subfigure[MRPs]{\label{fig:mrp2}
    \includegraphics[width=0.46\columnwidth]{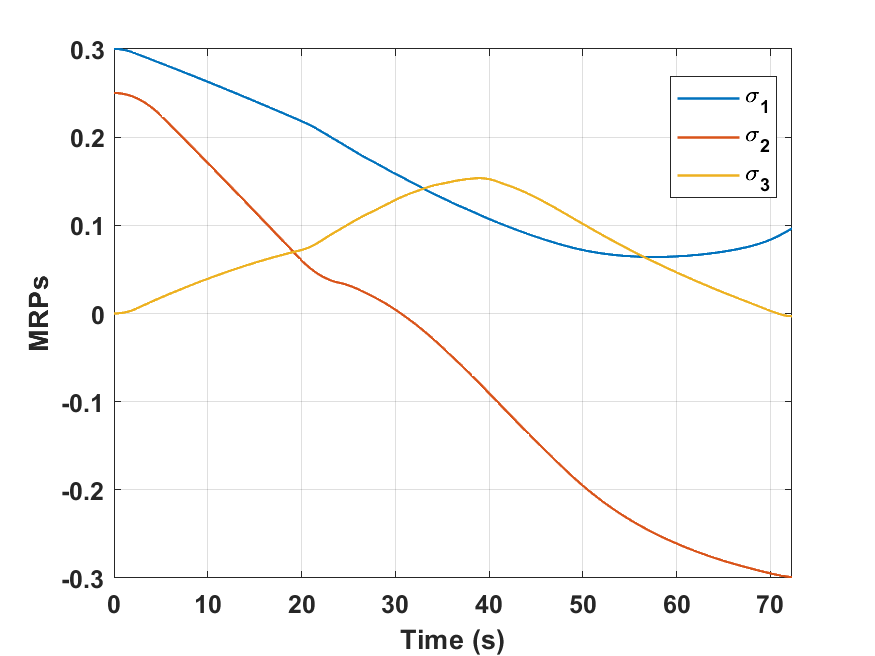}}
    \subfigure[Angular velocity]{\label{fig:angvel2}   
    \includegraphics[width=0.46\columnwidth]{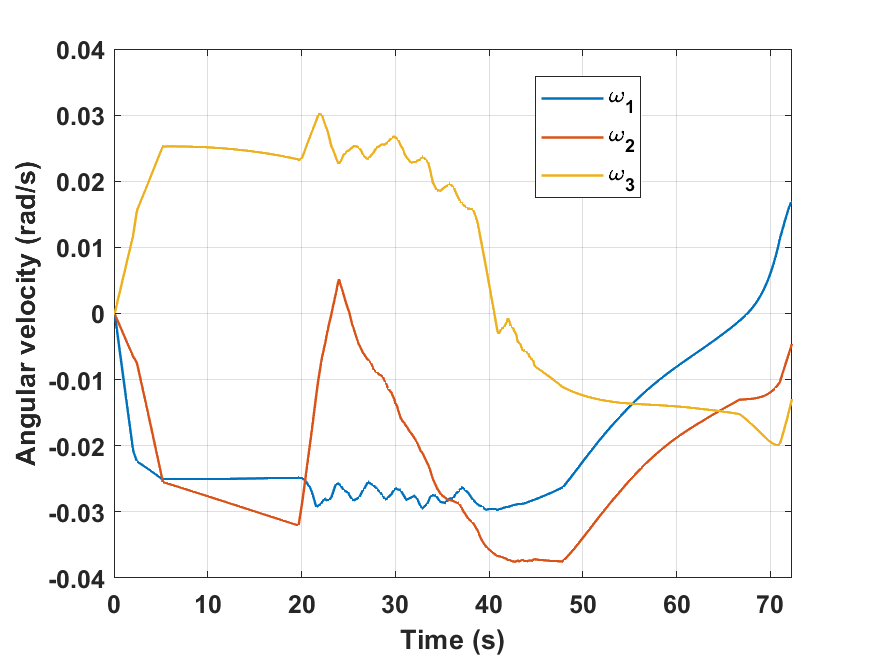}}
    \subfigure[Torque]{\label{fig:torque2}   
    \includegraphics[width=0.46\columnwidth]{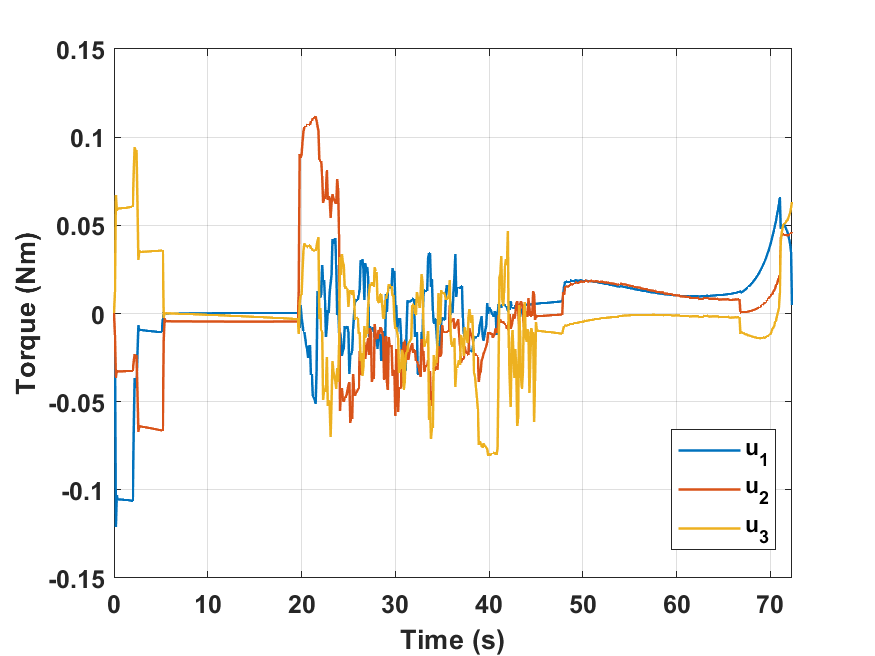}}
    \caption{Time History of the States (Case 2)}
    \label{fig:timehist2}
\end{figure}

\section{Conclusion}
This work proposes a constrained attitude maneuver using a path-planning approach while considering multiple keep-out constraints around spacecraft.
The modified Rodrigues parameters (MRPs) are used as the attitude representation, and the constraints (keep-out zones) are converted into the MRP space. To find the constraint-compliant MRP trajectory, the fuzzy enhanced potential field is applied, and the dynamic and kinematic equations are utilized to obtain the physics-compliant states. The performance of the proposed approach is validated via numerical simulations. The simulation results demonstrate that the proposed approach produces smooth MRP trajectories while avoiding the keep-out zones. However, the angular velocity and torque profiles contain undesired peaks and fluctuations without considering 
the terminal angular velocity. Future work will focus on improving the smoothness of the other states and addressing the terminal state conditions.

\bibliographystyle{AAS_publication}   % Number the references.
\bibliography{ConstrainedSAC}   % Use references.bib to resolve the labels.

@Article{EarthObservation,
AUTHOR = {Zhao, Qiang and Yu, Le and Du, Zhenrong and Peng, Dailiang and Hao, Pengyu and Zhang, Yongguang and Gong, Peng},
TITLE = {An Overview of the Applications of Earth Observation Satellite Data: Impacts and Future Trends},
JOURNAL = {Remote Sensing},
VOLUME = {14},
YEAR = {2022},
NUMBER = {8},
ARTICLE-NUMBER = {1863},
DOI = {10.3390/rs14081863}
}

@ARTICLE{ImagingSatellite,
  author={Lei, Jiakun and Meng, Tao and Wang, Kun and Wang, Weijia and Sun, Shujian and Wang, Lei},
  journal={IEEE Transactions on Aerospace and Electronic Systems}, 
  title={Adaptive Reduced-Attitude Control for Spacecraft Boresight Alignment with Safety Constraints and Accuracy Requirements}, 
  year={2024},
  volume={},
  number={},
  pages={1-19},
  doi={10.1109/TAES.2024.3373718}
}

@ARTICLE{PF2024_1,
  author={Duan, Chao and Hu, Qinglei and Yang, Haoyang and Wu, Huai-Ning},
  journal={IEEE Transactions on Industrial Electronics}, 
  title={Constrained Control of Underactuated Spacecraft Using Artificial Potentials}, 
  year={2024},
  volume={},
  number={},
  pages={1-10},
  doi={10.1109/TIE.2024.3366202}
}

@article{PF2024_2,
title = {Spacecraft attitude reorientation control method based on potential function under complex constraints},
journal = {Aerospace Science and Technology},
volume = {144},
pages = {108738},
year = {2024},
doi = {https://doi.org/10.1016/j.ast.2023.108738},
author = {Bing Hua and Jie He and Hong Zhang and Yunhua Wu and Zhiming Chen},
}

@article{PF2021_1,
author = {Yang, Juntang and Duan, Yisheng and Ben-Larbi, Mohamed Khalil and Stoll, Enrico},
title = {Potential Field-Based Sliding Surface Design and Its Application in Spacecraft Constrained Reorientation},
journal = {Journal of Guidance, Control, and Dynamics},
volume = {44},
number = {2},
pages = {399-409},
year = {2021},
doi = {10.2514/1.G005026},
}

@article{PF_disadv,
title = {Constrained attitude maneuvers on SO(3): Rotation space sampling, planning and low-level control},
journal = {Automatica},
volume = {112},
pages = {108659},
year = {2020},
doi = {https://doi.org/10.1016/j.automatica.2019.108659},
author = {Xiao Tan and Soulaimane Berkane and Dimos V. Dimarogonas}
}

@article{PF_disadv2,
author = {Hu, Qinglei and Chi, Biru and Akella, Maruthi R.},
title = {Anti-Unwinding Attitude Control of Spacecraft with Forbidden Pointing Constraints},
journal = {Journal of Guidance, Control, and Dynamics},
volume = {42},
number = {4},
pages = {822-835},
year = {2019},
doi = {10.2514/1.G003606}
}

@inproceedings{prob_roadmap,
author = {E. Feron and M. Dahleh and E. Frazzoli and R. Kornfeld},
title = {A randomized attitude slew planning algorithm for autonomous spacecraft},
booktitle = {AIAA Guidance, Navigation, and Control Conference and Exhibit},
chapter = {},
pages = {},
year = {2012},
doi = {10.2514/6.2001-4155}
}

@article{PP2022,
author = {Calaon, Riccardo and Schaub, Hanspeter},
title = {Constrained Attitude Maneuvering via Modified-Rodrigues-Parameter-Based Motion Planning Algorithms},
journal = {Journal of Spacecraft and Rockets},
volume = {59},
number = {4},
pages = {1342-1356},
year = {2022},
doi = {10.2514/1.A35294}
}

@article{PP2015,
author = {Kjellberg, Henri C. and Lightsey, E. Glenn},
title = {Discretized Quaternion Constrained Attitude Pathfinding},
journal = {Journal of Guidance, Control, and Dynamics},
volume = {39},
number = {3},
pages = {713-718},
year = {2016},
doi = {10.2514/1.G001063}
}

@article{PP2017,
author = {Tanygin, Sergei},
title = {Fast Autonomous Three-Axis Constrained Attitude Pathfinding and Visualization for Boresight Alignment},
journal = {Journal of Guidance, Control, and Dynamics},
volume = {40},
number = {2},
pages = {358-370},
year = {2017},
doi = {10.2514/1.G001801}
}

@article{PP2020,
title = {Constrained attitude maneuvers on SO(3): Rotation space sampling, planning and low-level control},
journal = {Automatica},
volume = {112},
pages = {108659},
year = {2020},
doi = {https://doi.org/10.1016/j.automatica.2019.108659},
author = {Xiao Tan and Soulaimane Berkane and Dimos V. Dimarogonas}
}

@article{Choi2021Robotica,
    title={Intelligent cooperative collision avoidance via fuzzy potential fields},
    journal={Robotica},
    publisher={Cambridge University Press},
    author={Choi, Daegyun and Chhabra, Anirudh and Kim, Donghoon},
    year={2021},
    pages={1–20},
    DOI={10.1017/S0263574721001454}
}

@Article{Choi2021AP,
    AUTHOR = {Choi, Daegyun and Kim, Donghoon and Lee, Kyuman},
    TITLE = {Enhanced Potential Field-Based Collision Avoidance in Cluttered Three-Dimensional Urban Environments},
    JOURNAL = {Applied Sciences},
    VOLUME = {11},
    YEAR = {2021},
    NUMBER = {22},
    ARTICLE-NUMBER = {11003},
    URL = {https://www.mdpi.com/2076-3417/11/22/11003},
    ISSN = {2076-3417},
    DOI = {10.3390/app112211003}
}

@book{Schaub2003analytical,
  title={Analytical Mechanics of Space Systems},
  author={Schaub, H. and Junkins, J.L.},
  isbn={9781563470547},
  lccn={2003007343},
  series={AIAA education series},
  year={2003},
  publisher={American Institute of Aeronautics and Astronautics}
}

@inproceedings{Choi2024RPO,
    author = {Choi, Daegyun and Kim, Donghoon},
    title = {Enhancing Spacecraft Relative Motion Control through FIS Interpretability},
    booktitle = {2024 North American Fuzzy Information Processing Society Conference},
    year = {2024} 
}

@inproceedings{Choi2020EPF2D,
    author = {Daegyun Choi and Kyuman Lee and Donghoon Kim},
    title = {Enhanced Potential Field-Based Collision Avoidance for Unmanned Aerial Vehicles in a Dynamic Environment},
    booktitle = {AIAA Scitech 2020 Forum},
    pages = {1-7},
    eprint = {https://arc.aiaa.org/doi/pdf/10.2514/6.2020-0487},
    publisher = {AIAA},
    year = {2020},
}

@article{Kjllberg2016,
author = {Kjellberg, Henri C. and Lightsey, E. Glenn},
title = {Discretized Quaternion Constrained Attitude Pathfinding},
journal = {Journal of Guidance, Control, and Dynamics},
volume = {39},
number = {3},
pages = {713-718},
year = {2016},
doi = {10.2514/1.G001063}
}

@article{Yang2021,
author = {Yang, Juntang and Duan, Yisheng and Ben-Larbi, Mohamed Khalil and Stoll, Enrico},
title = {Potential Field-Based Sliding Surface Design and Its Application in Spacecraft Constrained Reorientation},
journal = {Journal of Guidance, Control, and Dynamics},
volume = {44},
number = {2},
pages = {399-409},
year = {2021},
doi = {10.2514/1.G005026}
}

\end{document}